\documentclass{aa}  
\usepackage{graphicx}
\usepackage{txfonts}
\usepackage{natbib}
\usepackage{color}
\usepackage{txfonts}
\usepackage{bm} 
\usepackage[colorlinks=true,linkcolor=blue,citecolor=blue,urlcolor=blue]{hyperref}

\usepackage[normalem]{ulem}

\begin{document}

   \title{Beyond solar metallicity}

   \subtitle{How enhanced solid content in disks reshapes low‑mass planet torques}

\author{Zs. Regály, 
          \inst{1}\thanks{E-mail: \href{regaly@konkoly.hu}{regaly@konkoly.hu}}
          A. Németh
          \inst{2}
          }
\institute{
HUN-REN CSFK Konkoly Observatory, MTA Centre of Excellence, Konkoly Thege M. út 15-17, Budapest, 1121, Hungary
\and
ELTE Eötvös Loránd University, Institute of Physics and Astronomy, Department of Astronomy, \\
   H-1117 Budapest, Pázmány Péter sétány 1/A, Hungary}
             
\date{}
 
  \abstract
   {The migration of low‑mass planets ($M_\mathrm{p}\leq10~M_\oplus$) is tightly controlled by the torques exerted by both the gas and solids in their natal disks. While canonical models assume a solar solid‑to‑gas mass ratio ($\epsilon\simeq0.01$) and neglect the back-reaction of the solid component of the disk, recent work suggests that enhanced metallicity can radically alter these torques.}  
   {We quantify how elevated metallicities ($\epsilon=0.03$ and $\epsilon=0.1$) modify the gas and solid torques acting on an Earth‑mass planet, test widely used linear scaling prescriptions, and identify the regimes where solid back‑reaction becomes decisive.}
   {We performed global, two‑dimensional hydrodynamic simulations that (i) treat solid material as a pressureless fluid fully coupled to the gas through drag and (ii) include the reciprocal back‑reaction force. The low-mass planet was maintained on a fixed circular orbit, and thus we computed static torques. The Stokes number was varied from 0.01 to 10, and three surface‑density slopes ($p=0.5,~1.0, \mathrm{and}~1.5$) and three accretion efficiencies ($\eta=0\%,~10\%,\mathrm{and}~100\%$) were explored. Predicted torques, obtained by rescaling canonical $\epsilon=0.01$ results, were compared with direct simulations.}
   {Solid torques scale nearly linearly with $\epsilon$, but gas torques deviate by $50-100\%$ and can even reverse sign for $\mathrm{St}\leq1$ in $\epsilon=0.1$ disks. These discrepancies arise from strong, feedback‑driven, asymmetric gas perturbations in the co‑orbital region, which are amplified by rapid planetary accretion. Accurate total torques are recovered only for $\mathrm{St}\geq3$, independent of $\epsilon$ or $\eta$; for $\mathrm{St}\leq2$ the linear prescription systematically overestimates the magnitude, sometimes predicting the wrong sign.}
   {Solid back-reaction in high‑metallicity environments can dominate the migration torque budget of low‑mass planets. Simple metallicity rescalings are therefore unreliable for $\mathrm{St}\leq2$, implying that precise migration tracks -- particularly in metal‑rich disks -- require simulations that fully couple solid and gas dynamics. These results highlight metallicity as a key parameter in shaping the early orbital architecture of planetary systems.}


   \maketitle

\section{Introduction}

It is well established that planets undergo changes in their semimajor axes due to gravitational interactions with their natal protoplanetary disk. 
In particular, low-mass planets (with masses $\lesssim 10~M_\oplus$) are generally subject to inward migration in locally isothermal disks as a result of disk torques \citep{Tanaka+2002ApJ}. 
Consequently, planets that remain embedded in the disk during its lifetime are at risk of spiraling into their host star.

Several mechanisms that potentially counteract this potentially catastrophic inward migration have been proposed, including the concept of planetary migration traps \citep{Lyra+2008,Sandor+2011ApJ,GuileraSandor2017A&A,Pierens2024A&A...684A.199P} and localized temperature perturbations within the disk \citep{Lyra+2010ApJ}. 
For comprehensive reviews of the physical processes governing planetary migration, see, for example, \cite{KleyNelson2012ARA&A..50..211K}, \cite{RaymondMorbidelli2020plas.book..287R}, and \cite{Paardekooper2023ASPC..534..685P}.

In recent years, numerical simulations of planet--disk interactions have introduced a new paradigm, emphasizing the pivotal role of the disk’s solid component in altering -- and in some cases reversing -- the direction of planetary migration.
It has been demonstrated that a pronounced asymmetry in the spatial distribution of solids near the planet can generate an additional torque, which may reach magnitudes comparable to that of the gas torque \citep{BenitezPessah2018}. 
Moreover, planetary accretion can further enhance this solid-induced torque, thereby amplifying its impact on the planet’s orbital evolution \citep{Regaly2020, Chrenko2024A&A...690A..41C,Pierens2023MNRAS.520.3286P}.
When these findings are applied within a global disk model that incorporates viscous accretion and X-ray photoevaporation of the gaseous component, along with dust evolution processes, it becomes evident that the torque exerted by the solid phase can significantly influence migration behavior \citep{Guilera+2023}. 
In particular, such models reveal that planets forming near the water-ice line can undergo substantial outward migration driven by solid torques \citep{Guilera2025ApJ...986..199G}.

The back-reaction of solid material on the gas is often neglected in numerical simulations of planet--disk interactions. 
Nonetheless, a growing body of work has underscored the significance of this feedback mechanism \citep{Gonzalezetal2017, Dipierroetal2018, Gonzalezetal2018, HsiehLin2020}.
Our recent study has shown that back-reaction from the solid component can substantially perturb the gas distribution in the vicinity of the planet, to the extent that the gas torque becomes positive, even under locally isothermal conditions \citep{Regaly2025A&A...694A.279R}. 
In addition, a recent three-dimensional hydrodynamic simulation demonstrated that this solid back-reaction can excite buoyancy resonances in a globally isothermal disk that would otherwise lack such instabilities \citep{Chametla2025MNRAS.541..847C}.

In the canonical scenario, the solid-to-gas mass ratio in protoplanetary disks is approximately one percent (e.g., \citealp{WilliamsCieza2011}). 
It is commonly assumed that the torque exerted by the solid component scales linearly with the total solid mass in the disk (see, e.g., \citealt{BenitezPessah2018}). 
This assumption may break down when the  back-reaction from the solid component becomes sufficiently strong to perturb the gas dynamics. 
Such perturbations can significantly affect the evolution of the solid component itself, resulting in a nonlinear and coupled torque response.

\citet{HsiehLin2020} show that higher-metallicity disks can produce solid torques strong enough to alter the migration direction of low-mass planets or to induce stochastic migration behavior. 
The stochastic nature of migration is associated with the excitation of small-scale vortices produced in inviscid simulations.
Their work also accounted for the dynamical back-reaction of the solid component on the gas, ensuring a fully coupled treatment of solid--gas interactions.

We present a comparative analysis of solid and gas torque predictions derived from simulations in canonical-metallicity disks and numerical simulations in high-metallicity disks. 
Specifically, we contrast models that assume a canonical metallicity and solid back-reaction with simulations of high-metallicity disks in order to evaluate how metallicity and  solid back-reaction interplay to influence planetary migration.

\section{Hydrodynamic model}

We performed two-dimensional, global hydrodynamic simulations of planet-disk interactions using the {\small GFARGO2} code \citep{RegalyVorobyov2017,Regaly2020}, a GPU-enabled version of the {\small FARGO} code \citep{Masset2000}.
The disk consisted of gaseous and solid components, and an embedded planet.
The solid component of the disk was a pressureless fluid while the gas is locally isothermal.
The dynamics of the gaseous and solid components,  perturbed by the planet, are described by the continuity and Navier-Stokes equations.
The dynamics of the solid component are determined by the drag force of gas arising due to the velocity difference between the two disk components.
Emphasize that we also took into account the reaction force of solid component, namely the back-reaction of  the solid dynamics on the gas.
See \citet{Regaly2025A&A...694A.279R} for a complete presentation of the governing equations.
To compute the dynamics of the solid, we followed the method of \citet{Stoyanovskayaetal2017,Stoyanovskayaetal2018}, which was successfully tested in several works \citep{Pierensetal2019,Vorobyovetal2019,Regaly2025A&A...694A.279R}.

We studied the effect of the solid-to-gas mass ratio (metallicity of the disk), which is given by $\epsilon = 0.01,~0.03,~$ and $0.1$.
For the canonical metallicity, the solid-to-gas mass ratio was set to $\epsilon_0=0.01$.
We investigated three different power law for the initial gas surface density: $p=0.5,~1,~$ and $1.5$.
For simplicity the solid component has a fixed Stokes number ($0.01\leq\mathrm{St}<10$) throughout the disk.

For this investigation, the planet's mass was set to $1~M_\oplus$ and kept on a fixed circular orbit at $R=1$.
Three models of the strengths of planetary accretion were created, assuming zero, 10 and 100 percent efficiency.
Zero efficiency essentially means that there is no accretion. For moderate ($10\%$) and strong ($100\%$) accretion, the planetary Hill sphere would be emptied within 10 and 1 orbits, respectively; see details in \citet{Regaly2025A&A...694A.279R}.

The initial conditions of the surface densities and velocities of gaseous and solid component were identical to that of applied in \citet{Regaly2025A&A...694A.279R}.
The size and geometry of the numerical grid was also identical to our previous investigation.
The simulations were run for 200 planetary orbits.
We emphasize that, by that time, the torques arising from both the gas and solid components of the disk are saturated to constant values.

To investigate the effect of metallicity on the torques felt by the embedded planet, we calculated the torque of solid component, $\Gamma_\mathrm{d}^\mathrm{BR}$, and the gaseous component, $\Gamma_\mathrm{g}^\mathrm{BR}$ assuming that the solid-to-gas mass ratio is set by the $\epsilon$ and the canonical metallicity is set by $\epsilon_0=0.01$. 
For torque normalization, we used the absolute value of gas torque $|\Gamma_\mathrm{g}^\mathrm{NBR}|$ measured in models where the solid back-reaction was neglected.
In the above expressions BR and NBR stand for back-reacting and non-back-reacting models, respectively.

It is natural to assume that the contribution of the solid component to the total torque scales linearly with the solid mass fraction of the disk, i.e., the metallicity. 
When the back-reaction of solids on the gas is neglected, this scaling is applicable, as stated in \citet{BenitezPessah2018,Regaly2020} and adopted by \citet{Guilera+2023}. 
However, once solid-gas back-reaction is included, the gas torque must be modified accordingly, with the magnitude of the correction depending on the amount of solid material. 
A minimal closure is to assume that the back-reaction-induced modification to the gas dynamics, and hence to the gas torque, is also linear in metallicity.
Thus, to predict the solid torques for higher metallicity disk, the solid torque measured in canonical disk is simply multiplied by the metallicity increase $\Gamma^\mathrm{BR}_\mathrm{d}=\Gamma^\mathrm{BR,\epsilon_0}_\mathrm{d}(\epsilon/\epsilon_0)$.
However, for the predicted gas torque, first we took the effect of the back-reaction into account and then amplified it by the metallicity increase $\Gamma_\mathrm{g}^\mathrm{BR}=\Gamma_\mathrm{g}^\mathrm{NBR,\epsilon_0}+\left(\Gamma_\mathrm{g}^\mathrm{BR,\epsilon_0}-\Gamma_\mathrm{g}^\mathrm{NBR,\epsilon_0}\right)(\epsilon/\epsilon_0).$
The method used to calculate the torque in the numerical simulation was the same as the one described in \citet{Regaly2025A&A...694A.279R}.

\section{Results}

\begin{figure*}
\includegraphics[width=18.5cm]{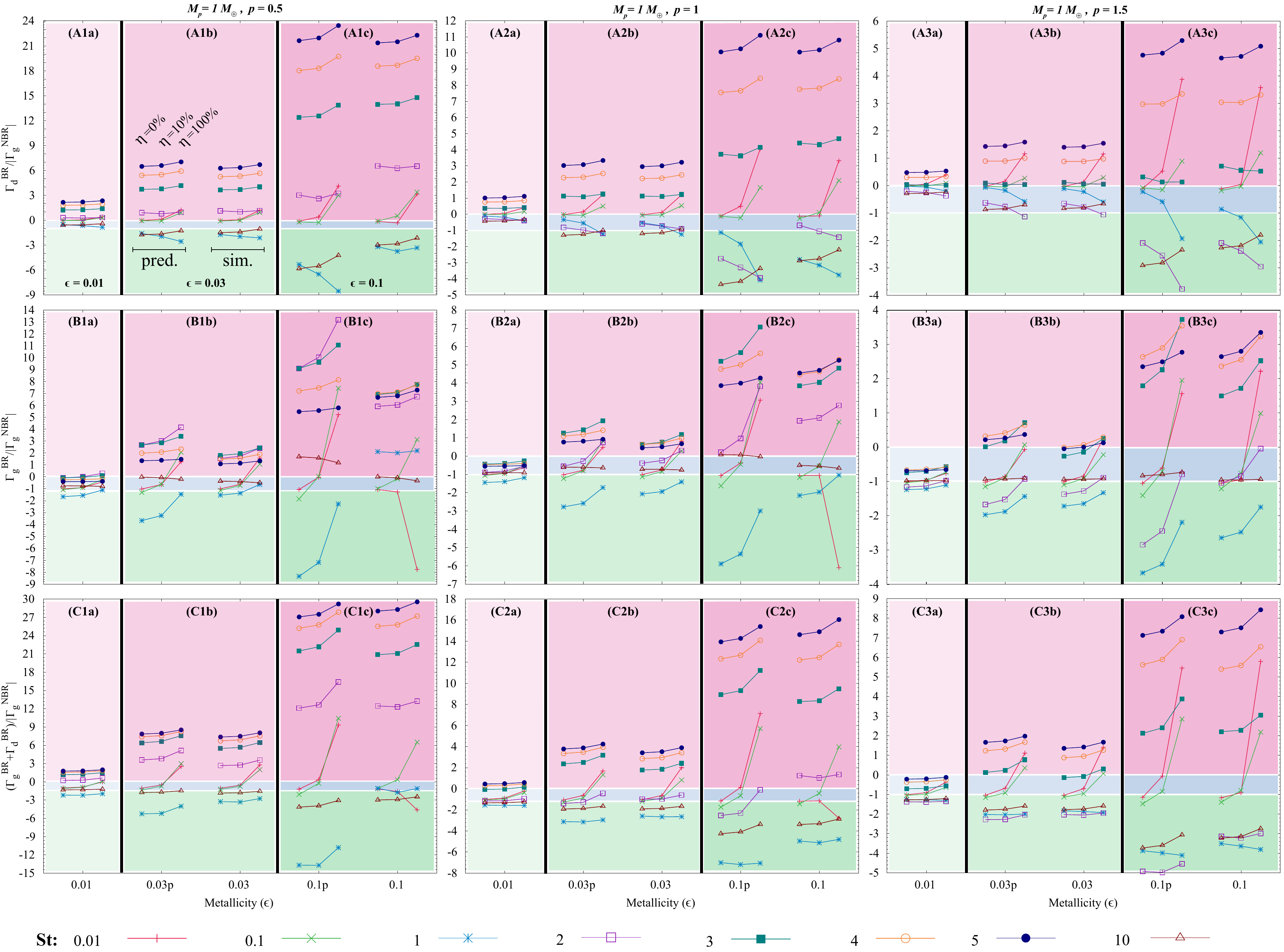}
\caption{Normalized torques felt by an$1~M_\oplus$ planet with metallicities of $\epsilon=0.01,~0.03,$ and $0.1$.
Three accretion efficiencies were modeled: $\eta= 0,~0.1,$ and 1.
The columns from left to right show three sets of models that assume different slopes for the initial disk density profile:$p = 0.5, 1.0,$ and 1.5.
Symbols represent the different Stokes numbers, $\mathrm{St}=0.01,~0.1,~1,~2,~3,~4,~5,$ and 10.
For the elevated metallicities, two models are shown: predicted (left) and numerical simulations (right).
The shaded regions of positive (red), weakened negative (blue), and strengthened negative (green) torques darken with the metallicity.
The normalization factor is the absolute value of the appropriate non-back-reacting gas torque.
Panels A1,~A2,~and A3: Solid torques in the back-reacting models normalized by the absolute value of the solid torque non-back-reacting models with $\epsilon=0.01$.
Panels B1,~B2,~and B3: Gas torques in the back-reacting models normalized by the absolute value of the gas torque in the non-back-reacting models with $\epsilon=0.01$.
Panels C1,~C2,~and C3: Total torques in the back-reacting models normalized by the absolute value of the gas torque in the non-back-reacting models with $\epsilon=0.01$.
}
\label{fig:tot_trq}
\end{figure*} 

Figure~\ref{fig:tot_trq} shows the solid, the gas, and the total torques felt by the $1~M_\oplus$ planet.
The three torque measurements connected by lines represent $\eta=0\% ,~10\%,~$ and $100\%$ accretion efficiencies.
The three metallicity models are highlighted using darker shades and are separated by black vertical lines.
In the horizontal axis, $p$ stands for the predicted models, while without it correspond to numerical simulations.
The three columns correspond to the three different ($p=0.5,~1,~$ and $1.5$) surface density slope.

Panels A1, A2, and A3 of Fig.~\ref{fig:tot_trq} show the normalized torque of the solid component, $\Gamma_\mathrm{d}^\mathrm{BR}/|\Gamma_\mathrm{g}^\mathrm{NBR}|$.
The subpanels, divided by the vertical black lines within each panel, represent the canonical, $\epsilon = 0.03$, and $\epsilon = 0.1$ metallicity cases, respectively.
First, let us discuss the canonical-metallicity model, $\epsilon=0.01$, shown with the lightest shade subpanels A1a, A2a, and A3a).
Note that there are no predicted torques shown for canonical model.
The torque of solid component usually positive for large Stokes numbers ($\mathrm{St}\geq2$).
As the initial density profile becomes steeper, the solid torques are reduced in magnitude.
Generally, the solid torque shows a weak dependence on the accretion efficiency for the canonical metallicity.

Panels B1, B2, and B3 of Fig.~\ref{fig:tot_trq} show the normalized gas torque, $\Gamma_\mathrm{g}^\mathrm{BR}/|\Gamma_\mathrm{g}^\mathrm{NBR}|$.
As can be seen, it is usually negative; however, it exhibits weakened magnitudes for the canonical metallicity (subpanels B1a, B2a, and B3a).
The degree of weakening decreases with an increasing steepness of the gas density profile.
As demonstrated previously in \citet{Regaly2025A&A...694A.279R}, the effect of solid back-reaction plays a significant role in shaping the gas torque magnitudes.

Panels C1, C2, and C3 of Fig.~\ref{fig:tot_trq} show the normalized total torque, ($\Gamma_\mathrm{g}^\mathrm{BR}+\Gamma_\mathrm{d}^\mathrm{BR})/|\Gamma_\mathrm{g}^\mathrm{NBR}|$.
In the canonical case (subpanels C1a, C2a, and C3a), the total torque can be positive for $p=0.5$ and large Stokes numbers, $2\leq\mathrm{St}\leq5$ or reduced magnitude negative (compared to the non-back-reacting models) for $p=1.0$ and $p=1.5$.
Note that no positive total torque was observed for $p=1.5$ models (subpanel C3a).

Now, let us compare the torque predictions and the results of the numerical simulations for three times the canonical metallicity, i.e., for a value of the parameter $\epsilon=0.03$.
The predicted torques of the solid components practically match those computed from the simulations (subpanels A1b, A2b, and A3b).
However, the predicted gas torques differ quantitatively from the predictions (subpanels B1b, B2b, and B3b).
Note that when making the prediction, we simply assumed that the effect of back-reaction was three times that observed in the canonical-metallicity model.
The largest discrepancy between the predicted gas torque and that computed from numerical simulations is about 50\% for $1\leq\mathrm{St}\leq5$.
For {the coupled species ($\mathrm{St}\leq0.1$), the discrepancy is modest.
Consequently, the total torques derived from numerical simulations can be accurately predicted for $\mathrm{St} \leq 0.1$ (see, e.g., $\mathrm{St}=0.01$, C1b).
However, for $1\leq\mathrm{St}\leq5$, the total torque magnitudes can be over-predicted by up to 100\%, as can be observed for $2\leq\mathrm{St}\leq3$ (e.g., subpanel C1b).
It should be emphasized that the above findings are not affected by accretion efficiency.

Now, let us focus on models that use ten times the canonical metallicity, i.e., with $\epsilon=0.1$.
Unlike the models with a metallicity of 0.03, in this case the predicted and numerical torques of the solid component show significant quantitative differences for $1\leq\mathrm{St}\leq3$.
The torques arising from these species are weaker or stronger in the numerical models than in the predictions, when they are negative or positive, respectively.
A much larger discrepancy is found with regard to the gas torque (subpanels B1c, B2c, and B3c).
For $\mathrm{St} = 0.01$ and $\mathrm{St} =1$, the sign of the torque can even mismatch.
For instance, in the $\mathrm{St} = 0.01$ and $p = 0.5$ or $p = 1.0$ models (see $\mathrm{St}=0.01$ in subpanels B1c and B2c), the torque predictions are positive for accreting planets with high accretion efficiency, but the numerical simulations give strong negative torques.
Another example is the $\mathrm{St} = 1$ and $p = 0.5$ model (subpanel B1c, $\mathrm{St}=1$), where the predictions give strong negative torques, while the numerical simulations result in significant positive torques, independent of the accretion efficiency.
Comparing the total torques (see the C1c, C2c, and C3c subpanels of Fig. \ref{fig:tot_trq}), we find that the predictions and numerical simulations agree well for $\mathrm{St}>2$, while significant discrepancies are found for $\mathrm{St} \leq 2$.
Moreover, the total torques arising in $\mathrm{St} = 0.01$ and $p = 0.5$ and $p = 1.0$ or $\mathrm{St} = 2$ and $p = 1.0$ models, can differ in sign due to the combined effect of changes in the solid and gas torques (see, e.g., $\mathrm{St}=2$ in subpanel C2c).

We now turn to models in which the predicted total torque shows qualitative discrepancies when compared to the outcomes of numerical simulations. 
These deviations are evident for solid particles with Stokes numbers $\mathrm{St} = 0.01$ and $\mathrm{St} = 1$.
As the torque contribution from the solid component is generally well captured by the analytical model, our attention is focused on the gas torque. 
Notably, qualitative mismatches between predicted and simulated total torques persist in these cases as well.

To understand the origin of these discrepancies, we analyzed the gas surface density perturbations induced by the planet in both canonical and high-metallicity disk simulations (see Sect.~\ref{sec:Disc} for details). 
Our analysis indicates that the qualitative disagreement between the predicted and simulated torques arises from a strongly asymmetric perturbation in the co-orbital region, with respect to the planet's orbital motion.

This asymmetry becomes most pronounced at the highest planetary accretion efficiency, $\eta = 100\%$, where the discrepancy between prediction and simulation is strongest. 

\section{Discussion}
\label{sec:Disc}

\begin{figure*}[ht!]
\includegraphics[width=0.9\textwidth]{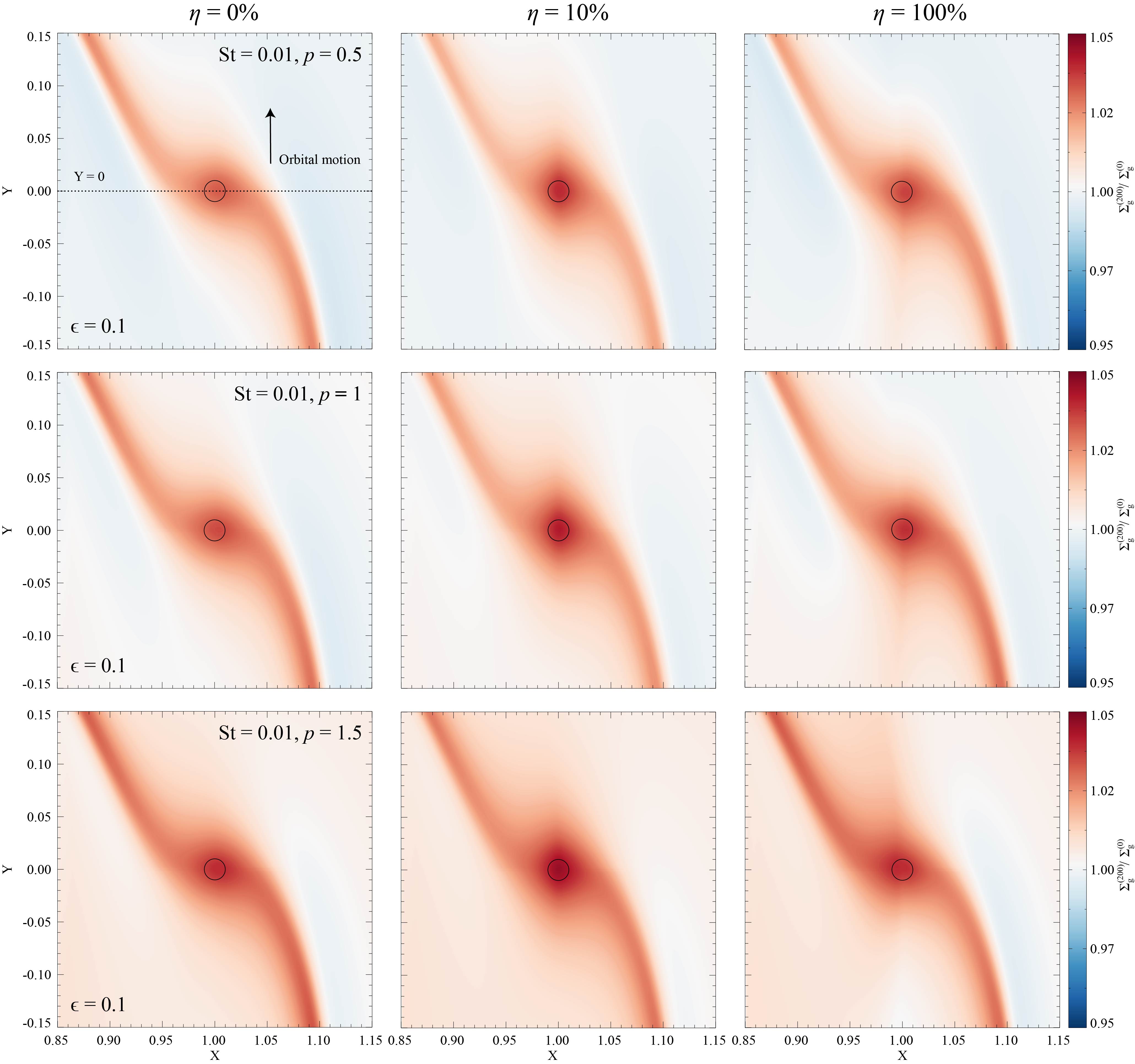}
\centering
\caption{Gas density distribution at the end of the simulation normalized with respect to the initial distribution around an Earth-mass planet for  $\mathrm{St}=0.01$ in a high-metallicity disk with $\epsilon=0.1$. The three different initial disk density profiles with steepness values of $p = 0.5, 1.0,$ and 1.5, and the three different accretion efficiencies with values of $\eta=0\%, 10\%,$ and $100\%,$ are shown. The circles represent the planetary Hill sphere. The orbital motion of the planet is indicated.}
\label{fig:distSt001}
\end{figure*}

\begin{figure*}[ht!]
\includegraphics[width=\textwidth]{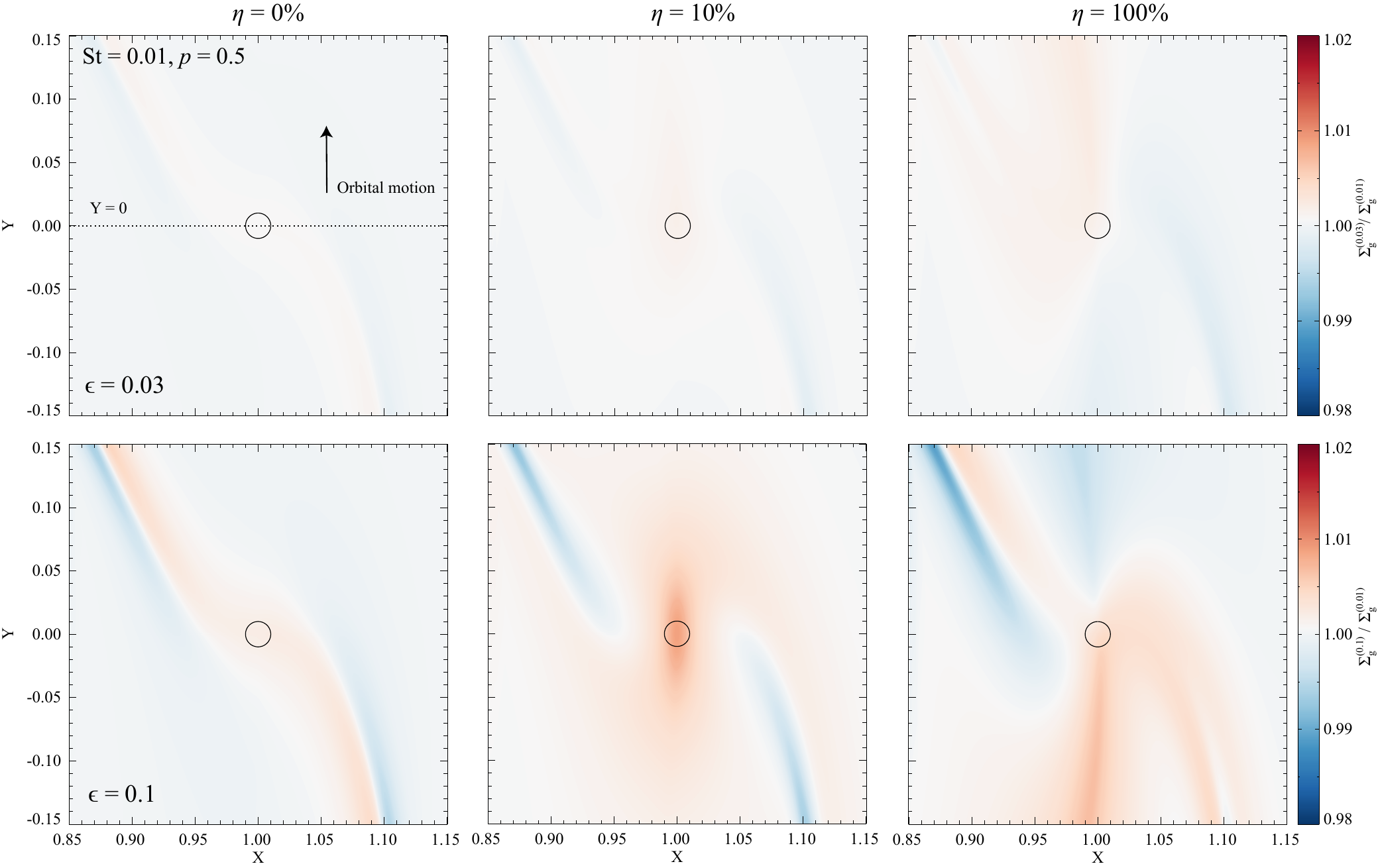}
\centering
\caption{Final gas density distribution normalized by that in the canonical-metallicity model in the vicinity of an Earth-mass planet, assuming a solid species with $\mathrm{St} = 0.01$ in a disk with a metallicity of $0.03$ (top) and $0.1$ (bottom) and a density profile of $p = 0.5$.}
\label{fig:diffSt001}
\centering
\end{figure*}

\begin{figure*}[ht!]
\includegraphics[width=\textwidth]{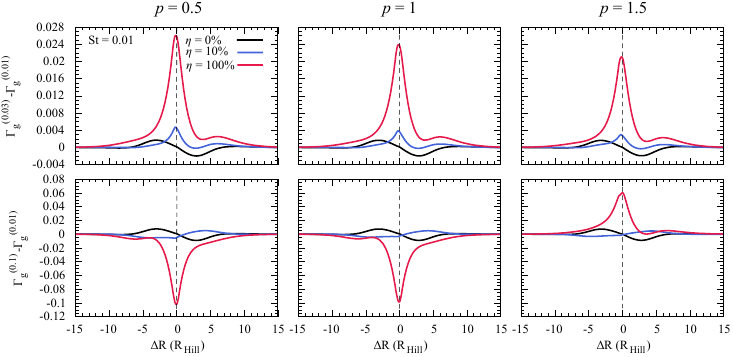}
\centering
\caption{Difference in the azimuthally averaged radial gas torque profiles for species with $\mathrm{St} = 0.01$ in disks of varying metallicities.
The different accretion efficiencies $\eta=0\%, 10\%$, and $100\%$ are shown in black, blue, and red, respectively.
}
\label{fig:torque_St001}
\end{figure*}

In this section we examine models in which the total torque prediction deviates not only quantitatively but also qualitatively from the outcomes of numerical simulations. Such discrepancies are found for solid species with $\mathrm{St} = 0.01$ and $\mathrm{St} = 1$.

Given that the torque exerted by the solid component of the disk is generally well captured by the prediction, our analysis focuses on the gas torque contribution. Notably, qualitative differences also emerge when comparing the total torques obtained from predictions and simulations.

Figure~\ref{fig:distSt001} displays the gas surface density distribution near the planet in the highest metallicity disk ($\epsilon = 0.1$) with $\mathrm{St}=0.01$. 
In this configuration, the theoretical model predicts a large positive total torque, whereas numerical simulations indicate a strong negative total torque in disks with $p = 0.5$ and $p = 1.0$ (see $\mathrm{St}=0.01$ in subpanels B1c and B2c of Fig. \ref{fig:tot_trq}). 
The gas snapshots, normalized by the initial gas density profile, correspond to the final state of the simulation. 
It is evident that the co-orbital region becomes strongly perturbed in the case of strong planetary accretion (rightmost column of Fig. \ref{fig:distSt001}). 
For moderate accretion (middle column), the co-orbital perturbation is still discernible. 
In the $p = 0.5$ and $p = 1.0$ disks (upper and middle rows), strong accretion leads to a gas density enhancement trailing the planet. 
In contrast, for the $p = 1.5$ disk, the density enhancement near the planet is symmetric with respect to the 
$Y=0$ axis, and an increase in gas density is observed interior to the planetary orbit.

Figure~\ref{fig:diffSt001} shows the gas surface density around the planet for $\mathrm{St}=0.01$ in $p=0.5$ disks with high metallicity, normalized to the corresponding canonical-metallicity distributions. 
At $\epsilon = 0.03$ (top panels), the gas density is only slightly enhanced ahead of the planet. 
At $\epsilon = 0.1$ (bottom panels), a more pronounced and symmetric enhancement develops around the $Y$-axis for non-accreting \textbf{($\eta = 0\%$)} and moderately accreting \textbf{($\eta = 10\%$)} planets. 
In the case of strong accretion \textbf{($\eta = 100\%$)}, however, a notable gas accumulation emerges behind the planet. 
This asymmetry is attributed to the enhanced back-reaction from the solid component of the disk, which significantly perturbs the local gas dynamics. 
A similar pattern is observed in $p = 1.0$ disks, while it is absent in the $p = 1.5$ models.

We quantified the deviation in gas torque between high- and canonical-metallicity disks in the planet's vicinity via
\begin{equation}
\Delta\Gamma^{(\epsilon=0.03)} = \Gamma^{(\epsilon=0.03)}_{\mathrm{g}} - \Gamma^{(\epsilon=0.01)}_{\mathrm{g}},
\end{equation}
and
\begin{equation}
\Delta\Gamma^{(\epsilon=0.1)} = \Gamma^{(\epsilon=0.1)}_{\mathrm{g}} - \Gamma^{(\epsilon=0.01)}_{\mathrm{g}}.
\end{equation}
Figure~\ref{fig:torque_St001} presents results for $\mathrm{St} = 0.01$ under varying planetary accretion efficiencies. 
For non-accreting or moderately accreting planets ($\eta = 0\%$ or $10\%$), the torque difference $|\Delta\Gamma^{(\epsilon=0.03)}|$ remains below 0.4\% (top panels). 
In contrast, for strongly accreting planets ($\eta = 1$), the deviation increases to 2.4\%. 
As indicated by Fig.~\ref{fig:diffSt001}, the structural differences in the gas density between canonical and higher metallicity disks are relatively minor, allowing for reasonably accurate qualitative predictions of gas torques.
However, for $\epsilon = 0.1$, $\Delta\Gamma^{(\epsilon=0.1)}$ becomes negative in strongly accreting cases within $p = 0.5$ and $p = 1.0$ disks (bottom panels of Fig. \ref{fig:torque_St001}). 
Interestingly, in the $p = 1.5$ disk, the torque difference turns positive under the same accretion conditions. 
This implies a qualitative mismatch between the predicted and simulated gas torques: while the prediction yields a strongly positive torque, simulations produce a strong negative torque. 
The origin of this discrepancy lies in the pronounced gas accumulation behind the planet in high-metallicity environments, as depicted in Fig. \ref{fig:diffSt001}.

\begin{figure*}[ht!]
\includegraphics[width=\textwidth]{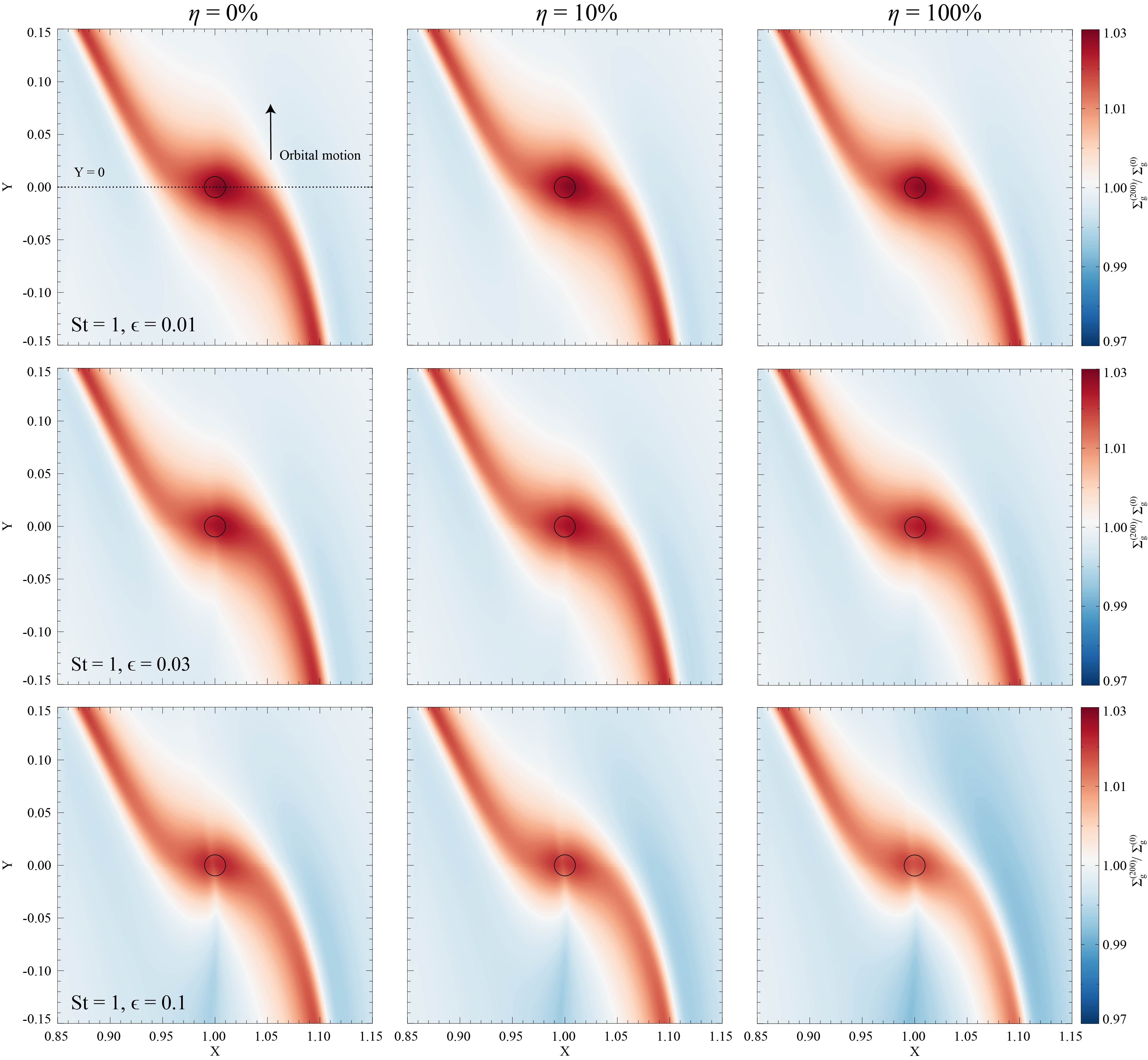}
\centering
\caption{Gas density distribution at the end of the simulation normalized with respect to the initial distribution around an Earth-mass planet for  $\mathrm{St}=1$ in  disk assuming metallicities of $\epsilon=0.01,~0.03,$ and $0.1$ and $p=0.5$. The three different accretion efficiencies, $\eta=0\%, 10\%,$ and $100\%,$ are shown. The circles represent the planetary Hill sphere. The orbital motion of the planet is indicated.}
\label{fig:distSt1}
\centering
\end{figure*}

\begin{figure*}[ht!]
\includegraphics[width=\textwidth]{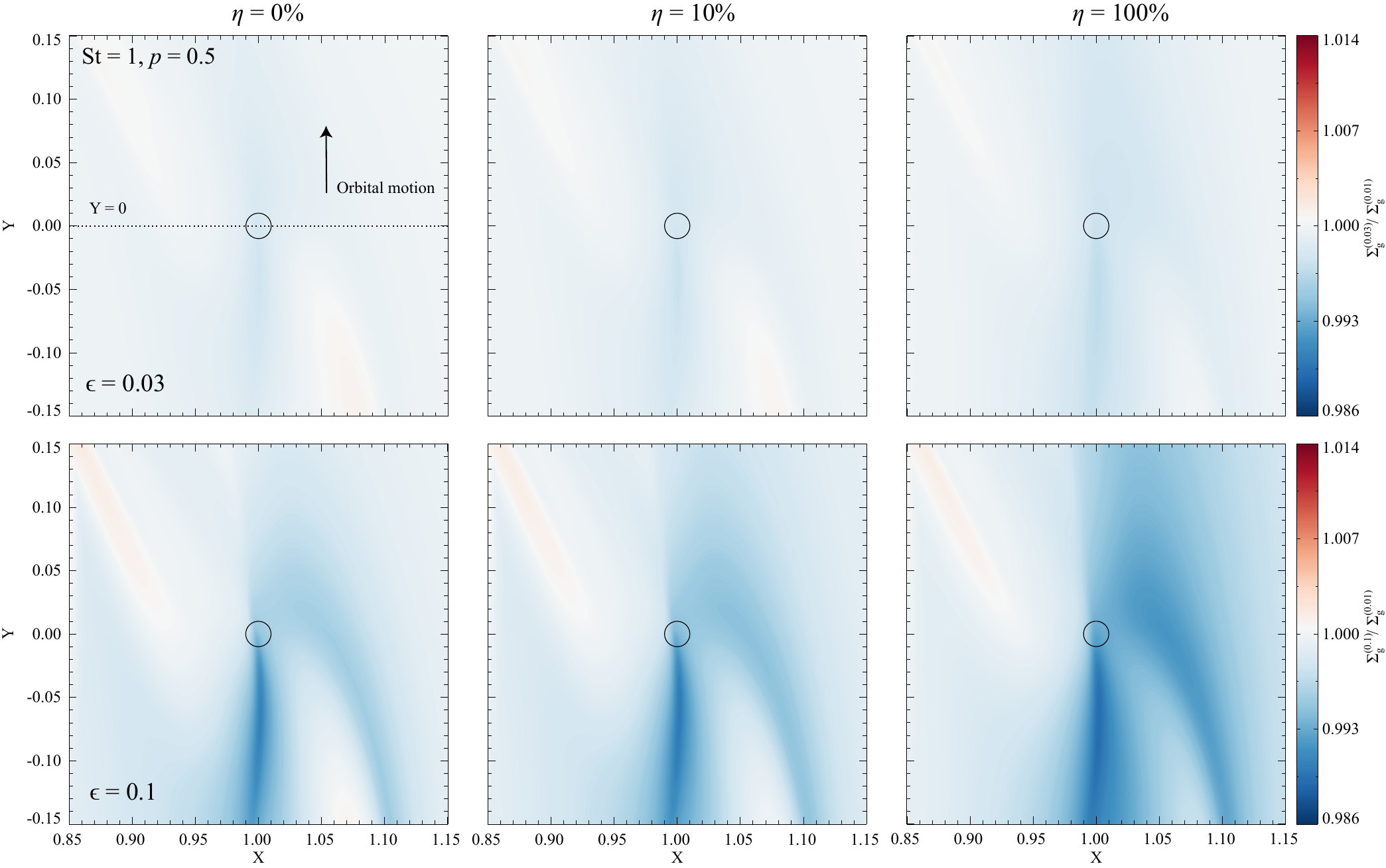}
\centering
\caption{Final gas density distribution normalized by that in the canonical-metallicity model in the vicinity of an Earth-mass planet, assuming a solid species with $\mathrm{St} = 1$ in a disk with a metallicity of $0.03$ (top) and $0.1$ (bottom) and a density profile of $p = 0.5$.}
\label{fig:diffSt1}
\centering
\end{figure*}

\begin{figure*}
\includegraphics[width=\textwidth]{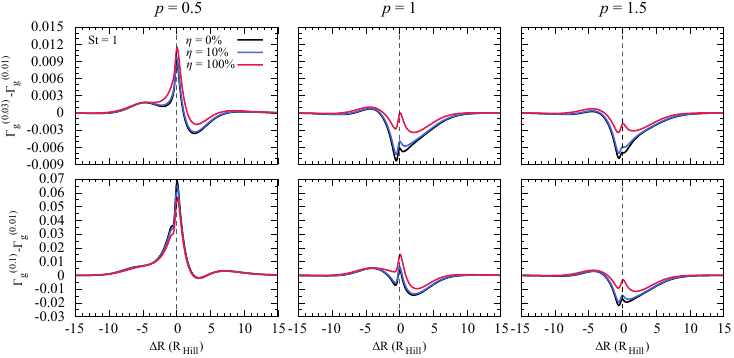}
\caption{Difference in the azimuthally averaged radial gas torque profiles for species with $\mathrm{St} = 1$ in disks of varying metallicities.
The different accretion efficiencies $\eta=0\%, 10\%$, and $100\%$ are shown in black, blue, and red, respectively.}
\label{fig:torque_St1}
\end{figure*}

We now turn our attention to cases where the prediction gives a strongly negative gas torque, yet the numerical simulations yield positive values for $\mathrm{St}=1$. 
This behavior is particularly evident in high-metallicity disks with $p = 0.5$ (see $\mathrm{St}=1$ in subpanel B1c of Fig. \ref{fig:tot_trq}).

Figure~\ref{fig:distSt1} presents the gas surface density distribution at the end of the simulation, in the vicinity of the planet embedded in a high-metallicity $p = 0.5$ disk with $\mathrm{St}=1$. 
The density is normalized by the initial gas profile. 
Similar to earlier cases, the co-orbital region is perturbed at elevated metallicities (middle and bottom panels). 
Notably, as the metallicity increases, the gas depletion behind the planet becomes progressively more pronounced.
This depletion in gas density trailing the planet causes the torque exerted by the gas to change sign and become positive when $\epsilon = 0.1$, regardless of the efficiency of accretion.

Figure~\ref{fig:diffSt1} shows the normalized gas surface density in the $p = 0.5$ disk for high metallicities, relative to the canonical-metallicity case for $\mathrm{St}=1$. 
At $\epsilon = 0.03$ (top panels), the gas distribution exhibits only slight enhancement along the planetary spiral waves, with a modest depression in the co-orbital region trailing the planet. 
However, at higher metallicity ($\epsilon = 0.1$, bottom panels), the trailing depletion becomes substantially deeper. 
Additionally, an asymmetry emerges in the spiral waves: the leading spiral wave shows a marked density enhancement, while the trailing wave exhibits a corresponding depletion. 
This asymmetry also contributes to the gas torque becoming positive in the high-metallicity ($\epsilon = 0.1$) case.
These features are characteristic signatures of the disk’s increased metallicity.

The resulting total torque differences for $\mathrm{St} = 1$ are illustrated in Fig.~\ref{fig:torque_St1}. 
The trend is consistent across all considered accretion efficiencies: in the $p = 0.5$ disk, the torque difference is positive (left column), while in the $p = 1.0$ and $p = 1.5$ disks, it remains negative (middle and right column). 
Although the gas density depletion behind the planet is significant, the corresponding positive gas torque is insufficient to fully counteract the negative torque contribution from the solid component at $\epsilon = 0.03$ (see $\mathrm{St}=1$ in subpanel C1b of Fig. \ref{fig:tot_trq}). 
In contrast, for the $p = 0.5$ disk at $\epsilon = 0.1$, the effect of back-reaction becomes dominant, leading to a substantial reduction in the total torque magnitude (see $\mathrm{St=1}$ in subpanel C1c of Fig. \ref{fig:tot_trq}).

Thus, for all three accretion scenarios modeled, the gas torque becomes strongly positive, contrary to the predictions. 
Since the solid torque of species with $\mathrm{St} = 1$ is negative and approximately half the predicted magnitude, the resulting total torque is weakly negative. 
This outcome contrasts sharply with predictions, which would suggest a strongly negative total torque in this regime.

\section{Caveats}

Here, we outline several caveats and suggest directions for future work aimed at improving the modeling of planetary migration in the low-mass regime. 
A more comprehensive discussion is also provided in \citet{Regaly2025A&A...694A.279R}.

Most importantly, the planet was held on a fixed orbit and thus only torques were computed. 
\citet{Paardekooper2014MNRAS.444.2031P} formalizes the distinction of static (migration-rate-independent) versus dynamical (migration-rate-dependent) components for low-mass planets and shows that dynamical corotation torques arise when the planet actually migrates -- these can strongly modify both the magnitude and sign relative to the static torque.
Numerous studies have demonstrated that planetary migration can alter the distribution and accumulation of solid material in the planet’s vicinity, thereby exerting a substantial impact on the surrounding gas dynamics (see, e.g., \citealp{Meru2019MNRAS.482.3678M,Wafflard-Fernandez2020MNRAS.493.5892W,Kanagawa2021ApJ...921..169K,Kuwahara2024A&A...692A..45K}). 
For instance, in a recent work \citet{Chametla2025A&A...698A..21C} showed that a dust-depleted region trailing the planet can drive either runaway migration or oscillatory torque-driven migration.

Another limitation of our study is that it focused exclusively on planets with a mass of $1~M_\oplus$, for which the influence of the solid component on the total torque is most pronounced \citep{Regaly2025A&A...694A.279R}. 
Exploring a broader range of planetary masses would provide insight into how solid back-reaction in high-metallicity disks scales with planetary mass.
Furthermore, modeling particles with large Stokes numbers ($\mathrm{St} > 1$) may require a Lagrangian treatment, as demonstrated by \citet{Chrenko2024A&A...690A..41C}.
Our simulations were performed in two dimensions; however, fully three-dimensional hydrodynamic models may uncover additional phenomena. For example, \citet{Chametla2025A&A...698A..21C} demonstrated that the structure of the dust-depleted region (dust-void) is significantly altered in three dimensions.
Finally, we employed a constant Stokes number approximation; this assumption may be inaccurate in regions with strong gas density perturbations, where the local gas properties -- and thus the effective Stokes number for a given particle size -- can vary significantly.
Future work should aim to address these simplifications to provide a more realistic and predictive framework for the study of low-mass planetary migration.

\citet{HsiehLin2020} found that small-scale vortices induced by the planet are excited only in inviscid, high-resolution simulations. 
When the disk viscosity was set to a value corresponding to $\alpha=3\times10^{-4}$, the vortices were found to be absent (see Sect. 3.4 in \citealp{HsiehLin2020}).
We also confirm that no vortices are excited in our simulations where the disk viscosity was modeled using the Shakura-Sunyaev $\alpha$-prescription with $\alpha=3\times10^{-4}$.
Furthermore, a key difference between our study and that of \citet{HsiehLin2020} is that we additionally included the diffusion of solids, as described by Eqs.~(3-4) and (12-13) in \citet{Regaly2025A&A...694A.279R}.

\section{Conclusions}

We investigated the effect of disk metallicity, modeled by changing the solid-to-gas mass ratio of the disk ($\epsilon$), on the torque exerted on a $1~M_\oplus$ planet by the disk's solid and gaseous components.
The planet was maintained on a fixed circular orbit, and consequently, only static torques were computed. These may differ from the torques that would arise in a fully dynamical (migrating) configuration.
We investigated higher disk metallicities in the cases of $\epsilon=0.03$ and $\epsilon=0.1$ and compared them to the canonical case of $\epsilon = 0.01$.
Based on our earlier models \citep{Regaly2025A&A...694A.279R}, the torque components were predicted by considering both the torque exerted by the solid component and its back-reaction on the gas dynamics.
To verify these predictions, we conducted two-dimensional numerical hydrodynamic simulations of planet--disk interactions in high-metallicity disks.
In these simulations, we took the back-reaction of the solid disk component on the gas dynamics into account.
Our main findings are as follows:

\begin{enumerate}

\item 
The torque contribution from the solid component of the disk is qualitatively well reproduced by theoretical predictions. In contrast, the gas torque often exhibits substantial quantitative discrepancies -- ranging from 50\% to 100\% -- and, in some cases, even qualitative differences, such as a reversal in sign.
Consequently, the total torque experienced by a $1~M_\oplus$ planet, arising from both gas and solid components, can only be reliably predicted in high-metallicity disks for particles with $\mathrm{St} \geq 3$, irrespective of the planetary accretion efficiency.

\item 
The predictions generally overestimate the total torque amplitude for $\mathrm{St} \leq 2$. These overestimates are most prominent in models with the highest metallicity ($\epsilon=0.1$) and the highest accretion efficiency ($\eta=100\%$).

\item For species with $\mathrm{St}=1$, the predictions yield a very strong negative torque. However, the numerical simulations yield a total torque magnitude that is an order of magnitude smaller ($p=0.5$) or half the magnitude ($p=1.0$).
This phenomenon also occurs  for solid species with $\mathrm{St} = 2$ in disks with a steep surface density profile ($p = 1.5$).

\item 
For coupled solid species with $\mathrm{St}=0.01$, numerical simulations result in a negative total torque when planetary accretion is significant ($10\%$ or $100\%$) and the surface density slope is shallow ($p = 0.5$ or $1.0$). However, the predictions suggest a strong positive total torque.

\end{enumerate}

Previously we showed that the gas torque is significantly altered due to the back-reaction of solid species \citep{Regaly2025A&A...694A.279R}.
Since the effect of the back-reaction of solid species is enhanced in high-metallicity disks, the resulting change in the gas torque significantly alters the total torque felt by the planet.
Thus, care must be taken when extrapolating the total torque from canonical-metallicity ($\epsilon=0.01$) disk simulations.
The simplified assumptions that the torque is simply multiplied by the elevated solid-to-gas mass ratio is not valid for all Stokes number investigated ($0.01\leq\mathrm{St}\leq10$).
In light of our findings, migration studies of low-mass planets ($0.1~M_\oplus\leq M_\mathrm{p}\leq10~M_\oplus$) require numerical simulations that use proper solid content in the disk instead of predicting the torque from canonical-metallicity disk models.

\begin{acknowledgements}
     We thank the anonymous referee, whose helpful comments  significantly helped improve the quality of this paper.
\end{acknowledgements}

\bibliographystyle{aa}
\bibliography{ref}

\appendix

\end{document}